\documentclass[letterpaper,english,conference]{IEEEtran}
\usepackage{setspace,mathdots,comment} 

\usepackage[usenames]{color}
\usepackage{prettyref}
\usepackage{graphicx,amsmath,amssymb,enumerate}

\usepackage{amsmath,amssymb,amsthm,graphicx,graphics,subfigure,epsfig,enumerate,bm}
\usepackage{ccaption}

\usepackage{algorithm}
\usepackage{algorithmic}

\newcommand{\bx}{{\boldsymbol x}}

\newcommand{\by}{{\boldsymbol y}}

\newcommand{\argmax}{\mathop{\rm argmax}}
\newcommand{\argmin}{\mathop{\rm argmin}}

\newcommand{\bc}{{\boldsymbol c}}

\newcommand{\bu}{{\boldsymbol u}}
\newcommand{\bg}{{\boldsymbol g}}
\newcommand{\bv}{{\boldsymbol v}}
\newcommand{\bI}{{\boldsymbol I}}

\newcommand{\bW}{{\boldsymbol W}}

\newrefformat{eq}{(\ref{#1})}
\newrefformat{sec}{Section~\ref{#1}}
\newrefformat{algo}{Algorithm~\ref{#1}}
\newrefformat{fig}{Fig.~\ref{#1}}
\newrefformat{tab}{Table~\ref{#1}}
\newrefformat{rmk}{Remark~\ref{#1}}
\newrefformat{clm}{Claim~\ref{#1}}
\newrefformat{def}{Definition~\ref{#1}}
\newrefformat{cor}{Corollary~\ref{#1}}
\newrefformat{lmm}{Lemma~\ref{#1}}
\newrefformat{lemma}{Lemma~\ref{#1}}
\newrefformat{prop}{Proposition~\ref{#1}}

\newtheorem{theorem}{Theorem}

\usepackage{algorithm, multirow}
\usepackage{algorithmic}
\usepackage{graphicx}
\usepackage{amsmath}
\usepackage{amssymb}
\usepackage{wrapfig}

\DeclareMathOperator{\trace}{Tr}
\DeclareMathOperator{\toep}{toep}

\newcommand{\cA}{{\mathcal A}}

\theoremstyle{definition}

\begin{document}
\theoremstyle{plain}\newtheorem{lemma}{\textbf{Lemma}}\newtheorem{corollary}{\textbf{Corollary}}\newtheorem{assumption}{\textbf{Assumption}}\newtheorem{prop}{\textbf{Proposition}}
\newtheorem{example}{\textbf{Example}}
\theoremstyle{remark}\newtheorem{remark}{\textbf{Remark}}
%
\title{Super-Resolution of Mutually Interfering Signals}

\author{\IEEEauthorblockN{Yuanxin Li}
\IEEEauthorblockA{Department of Electrical and Computer Engineering \\
The Ohio State University\\
Columbus, Ohio, 43210\\
Email: li.3822@osu.edu}
\and
\IEEEauthorblockN{Yuejie Chi}
\IEEEauthorblockA{Department of Electrical and Computer Engineering\\
The Ohio State University\\
Columbus, Ohio, 43210\\
Email: chi.97@osu.edu}
}


%


\maketitle

\begin{abstract}
We consider simultaneously identifying the membership and locations of point sources that are convolved with different low-pass point spread functions, from the observation of their superpositions. This problem arises in three-dimensional super-resolution single-molecule imaging, neural spike sorting, multi-user channel identification, among others. We propose a novel algorithm, based on convex programming, and establish its near-optimal performance guarantee for exact recovery by exploiting the sparsity of the point source model as well as incoherence between the point spread functions. Numerical examples are provided to demonstrate the effectiveness of the proposed approach.

\end{abstract}



%

\section{Introduction}
In many emerging applications in applied science and engineering, the acquired signal at the sensor can be regarded as a superposition of returns from multiple channels (or users), where the return from each channel is governed by the underlying physical field that produced it, e.g. the Green's function, the point spread function, the signature waveform, etc. The goal is to {\em invert} for the field parameters of each channel that produced the acquired signal which reflects the ensemble behavior of all channels. 

Mathematically, consider the acquired signal, $y(t)$, given as
\begin{equation}\label{demixing}
y(t) = \sum_{i=1}^I x_i( t )* g_i( t) = \sum_{i=1}^I \left( \sum_{k=1}^{K_i} a_{ik}g_i(t- \tau_{ik}) \right), 
\end{equation}
where $*$ denotes convolution, $x_i(t) = \sum_{k=1}^{K_i}a_{ik}\delta(t-\tau_{ik})$ is the point source signal observed through the $i$th channel, $g_i(t)$ is the point spread function of the $i$th channel, respectively, and $I$ denotes the total number of channels. For the $i$th channel, let ${\tau}_{ik}\in [0,1)$ and $a_{ik}\in\mathbb{C}$ be the location and the amplitude of the $k$th point source, $1\leq k\leq K_i$, respectively. In typical applications, we are interested in resolving the point sources of each channel at a resolution much higher than that of the acquired signal $y(t)$, determined by the Rayleigh limit, or in other words, the bandwidth of the point spread functions. 

The proposed model \eqref{demixing} occurs in a wide range of practical problems, ranging from three-dimensional super-resolution single-molecule imaging, to spike sorting in neural recording and DNA sequencing, to multi-user multi-path channel identification in communication systems, and many others.

\noindent\textbf{Three-dimensional super-resolution imaging:} By employing photoswitchable fluorescent molecules, the imaging process of stochastic optical reconstruction microscopy (STORM) \cite{rust2006sub} is divided into many frames, where in each frame, a {\em sparse} number of fluorophores (point sources) are randomly activated, localized at a resolution below the diffraction limit, and deactivated. The final image is thus obtained by superimposing the localization outcomes of all the frames. This principle can be extended to reconstruct a 3-D object from 2-D image frames \cite{huang2008three}, by introducing a cylindrical lens to modulate the ellipticity of the point spread function based on the depth of the fluorescent object. Therefore, the acquired image in each frame can be regarded as a {\em superposition} of returns from multiple depth layers, where the return from each layer corresponds to the convolution outcome of the point sources in that depth layer with the depth-dependent PSF, as modeled in \eqref{demixing}. The goal is thus to recover the locations and depth membership of each point source given the image frame that records the returns from all depth layers. 

\noindent\textbf{Spike sorting for neural recording:} Neurons in the brain communicate by firing action potentials, i.e. spikes, and it is possible to hear their communications through a single or multiple microelectrodes, which record simultaneously activities of multiple neurons within a local neighborhood. Spike sorting \cite{lewicki1998review}, thus, refers to the grouping of spikes according to each neuron, from the recording of the microelectrodes. Interestingly, it is possible to model the spike fired by each neuron with a {\em characteristic} shape \cite{gerstein1964simultaneous}. The neural recording can thus be modeled as a {\em superposition} of returns from multiple neurons, as in \eqref{demixing}, where the return of each neuron corresponds to the convolution of its characteristic spike shape with the sequence of its firing times. A similar formulation also arises in DNA sequencing \cite{li2000parametric}.



\noindent\textbf{Multi-path identification in random-access channels:} In multi-user multiple access model \cite{applebaum2012asynchronous}, the base station receives a {\em superposition} of returns from active users, as in \eqref{demixing}, where the received signal component for each active user corresponds to the convolution of its signature waveform with the unknown {\em sparse} multi-path channel from the user to the base station. The goal is to identify the set of active users, as well as their channel states, from the received signal at the base station.

\subsection{Related Work and Contributions}

There's extensive research literature on inverting \eqref{demixing} when there is only a single channel with $I=1$, where conventional methods such as matched filtering, MUSIC \cite{Schmidt1986MUSIC}, matrix pencil \cite{HuaSarkar1990}, to more recent approaches based on total variation minimization \cite{candes2014towards}, can be applied. However, these approaches can not be applied directly when multiple channels exist in the observed signal, due to the mutual interference between channels. To the best of the authors' knowledge, methods for inverting \eqref{demixing} with multiple channels have been extremely limited. In \cite{applebaum2012asynchronous,romberg2010sparse,chi2013compressive}, sparse recovery algorithms have been proposed to invert \eqref{demixing} by assuming the source locations lie on a fine-grain grid, which is mismatched from the actual source locations, resulting in possibly performance degradation \cite{Chi2011sensitivity}. Even when all the point sources indeed lie on the grid, existing work \cite{chi2013compressive} shows that the sample complexity needs to grow logarithmically with the size of the discretized grid, which is undesirable. The continuous basis pursuit algorithm \cite{ekanadham2011recovery} can be applied to retrieve source locations off the grid, but lacks performance guarantees.


In this paper, we study the problem of super-resolving \eqref{demixing} when there're two channels, i.e. $I=2$. We start by recognizing that in the frequency domain, the observed signal can be regarded as a linear combination of two spectrally-sparse signals, each composed of a small number of complex sinusoids. We then separate and recover the two signals by motivating their spectral structures using atomic norm minimization, which has been recently shown as an efficient convex optimization framework to motivate parsimonious structures \cite{Chandrasekaran2010convex,candes2014towards,tang2014CSoffgrid}, as well as satisfying the observation constraints. 

The separation and identification of the two channels, using the proposed algorithm, denoted by {\em convex demixing}, is made possible with two additional conditions. The first condition is that the point source models satisfy a mild separation condition, such that the locations of the point sources are separated by at least four times the Rayleigh limit; this is in line with the separation condition required by Cand\`es and Fernandez-Granda \cite{candes2014towards} even with a single channel. The second condition is that the point spread functions of different channels has to be sufficiently incoherent for separation, which is supplied in our theoretical analysis by assuming they're randomly generated from a uniform distribution on the complex circle. We demonstrate that, as soon as the number of measurements, or alternatively, the bandwidth of the point spread functions, is on the order $\max(K_1,K_2)\log(K_1+K_2)$ up to logarithmic factors, the proposed algorithm, denoted by {\em convex demixing}, recovers the locations of the point sources for each channel exactly, with high probability. Since at least an order of $K_1+K_2$ measurements is necessary, our sample complexity is near-optimal up to logarithmic factors. Moreover, the point sources can be recovered from the dual solution of the proposed algorithm, without estimating or knowing the model order a priori. Our proof is based on constructing two related polynomials that certify the optimality of the proposed algorithm. The effectiveness of the proposed algorithm is demonstrated in numerical experiments.

\subsection{Organization}

The rest of this paper is organized as follows. The proposed algorithm based on convex programming and its performance guarantee are provided in Section~\ref{sec::main_result}. The proof of the performance guarantee is sketched in Section~\ref{sec::proof_sketch}. Numerical experiments are shown in Section~\ref{sec::numerical_experiment}, and we conclude the paper in Section~\ref{sec::conclusion}. Throughout the paper, $\left(\cdot\right)^{T}$ and $\left(\cdot\right)^{*}$ denote the transpose and Hermitian transpose respectively, $\bar{\left(\cdot\right)}$ denotes the element-wise conjugate, $\trace\left(\cdot\right)$ denotes the trace of a matrix, and $\toep\left(\bu\right)$ denotes a Toeplitz Hermitian matrix with $\bu$ as its first column.

\section{Super-resolution of Mutually Interfering Signals}\label{sec::main_result}
Denote the discrete-time Fourier transform (DTFT) of $g_i(t)$ as
\begin{equation}
g_{i,n} = \int_{-\infty}^{\infty} g_i(t)e^{-j2\pi nt}dt,
\end{equation}
which satisfies $g_{i,n} = 0$ whenever $n\notin\Omega_M=\left\{-2M,\dots,0,\dots,2M\right\}$, where $2M$ is the cut-off frequency of band-limited $g_i\left(t\right)$. Typically, $M$ is determined by the physics, such as the aperture of the imaging device or the steering array. Taking the DTFT of \eqref{demixing} with $I=2$, we obtain
\begin{align}\label{demixing_freq}
y_n & = \int_{-\infty}^{\infty} y(t)e^{-j2\pi nt}dt \nonumber \\
& = g_{1,n}\cdot\left(\sum_{k=1}^{K_1}a_{1k}e^{-j2\pi n\tau_{1k}}\right) +g_{2,n}\cdot\left(\sum_{k=1}^{K_{2}}a_{2k}e^{-j2\pi n\tau_{2k}}\right),
\end{align}
with $n\in\Omega_M$. The measurements $y_{n}$'s in \eqref{demixing_freq} can be considered as a linear combination of two spectrally-sparse signals, with $g_{in}$'s determining the combination coefficients. Multiplying both sides of \eqref{demixing_freq} with $g_{1n}^{-1}$, and with slight abuse of notation, we rewrite \eqref{demixing_freq} into a vector form:
\begin{equation}\label{measurement}
\by =   \bx_1^{\star} + \boldsymbol{g} \odot \bx_2^{\star},
\end{equation}
where $\by=\left[y_{-2M},\dots,y_{0},\dots,y_{2M}\right]^{T}\in\mathbb{C}^{4M+1}$, $\bg=\left[g_{-2M},\dots,g_{0},\dots,g_{2M}\right]^{T}\in\mathbb{C}^{4M+1}$ with $g_{n}=g_{2,n}/g_{1,n}$, and $\odot$ denotes the Hadamard element-wise product operator. Furthermore, let $\bx_{1}^{\star}=\left[x_{1,-2M},\dots,x_{1,0},\dots,x_{1,2M}\right]^{T}\in\mathbb{C}^{4M+1}$ and $\bx_{2}^{\star}=\left[x_{2,-2M},\dots,x_{2,0},\dots,x_{2,2M}\right]^{T}\in\mathbb{C}^{4M+1}$ denote two spectrally-sparse signals, each composed of a small number of distinct complex harmonics, represented as 
\begin{equation} 
\begin{split}
&\bx_{1}^{\star}=\sum_{k=1}^{K_{1}}a_{1k}\bc\left(\tau_{1k}\right)\in\mathbb{C}^{4M+1}, \\
&\bx_{2}^{\star}=\sum_{k=1}^{K_{2}}a_{2k}\bc\left(\tau_{2k}\right)\in\mathbb{C}^{4M+1},
\end{split}
\end{equation}
where $K_{1}$ and $K_{2}$ are the spectral sparsity levels of each signal. The atom $\bc\left(\tau\right)$ is defined as 
\begin{equation*}
\bc(\tau) = \left[e^{-j2\pi(-2M)\tau},\ldots, 1,\ldots, e^{-j2\pi(2M)\tau}\right]^T,
\end{equation*}
which corresponds to a point source at the location $\tau\in[0,1]$. Further denote the sets of point sources in $\bx_{1}^{\star}$ and $\bx_{2}^{\star}$ by $\Upsilon_{1}=\left\{\tau_{11},\dots,\tau_{1K_{1}}\right\}$ and $\Upsilon_{2}=\left\{\tau_{21},\dots,\tau_{2K_{2}}\right\}$ respectively. The goal is thus to recover $\Upsilon_1$ and $\Upsilon_2$, and their corresponding amplitudes, from \eqref{measurement}.

Define the atomic norm \cite{Chandrasekaran2010convex,tang2014CSoffgrid} of $\boldsymbol{x}\in\mathbb{C}^{4M+1}$ with respect to the atoms $\bc(\tau)$ as 
\begin{equation*}
\left\Vert\bx\right\Vert_{\mathcal{A}}=\inf_{a_{k}\in\mathbb{C},\tau_{k}\in[0,1)}\left\{\sum_{k}\left\vert a_{k}\right\vert \mid \bx=\sum_{k} a_{k} \bc\left(\tau_{k}\right) \right\},
\end{equation*}
which can be regarded as the tightest convex relaxation of counting the smallest number of atoms that is needed to represent a signal $\boldsymbol{x}$. Therefore, we seek to recover the signals $\bx_1$ and $\bx_2$ by motivating their spectral sparsity via minimizing the atomic norm, with respect to the observation constraint: 
\begin{equation}\label{convex_demixing_rewritten}
\begin{split}
&\{\hat{\bx}_1, \hat{\bx}_2\} = \argmin_{\bx_1,\bx_2} \| \bx_1\|_{\cA} +  \|\bx_2\|_{\cA},\\
& \quad \quad \quad \quad \quad \mbox{s.t.} \quad \by= \bx_1 + \boldsymbol{g} \odot \bx_2.
\end{split}
\end{equation}
The above algorithm is referred to as {\em convex demixing}. Interestingly, \eqref{convex_demixing_rewritten} can be equivalently rewritten with the following semidefinite programming characterization \cite{tang2014CSoffgrid}, which can be solved efficiently using off-the-shelf solvers, as
\begin{equation}\label{convex_demixing_sdp}
\begin{split}
&\min_{\bx_1,\bx_2, \bu_{1}, \bu_{2}, t_{1}, t_{2}} \frac{\trace\left(\toep\left(\bu_{1}\right)\right) +  \trace\left(\toep\left(\bu_{2}\right)\right)}{4M+1}  + (t_1+t_{2}),\\
&\quad \quad \ \   \mbox{s.t.} \quad\ \  \begin{bmatrix}\toep\left(\bu_{1}\right)&\bx_{1}\\ \bx_{1}^{*}&t_{1}\end{bmatrix}\succeq \boldsymbol{0},\; \begin{bmatrix}\toep\left(\bu_{2}\right)&\bx_{2}\\ \bx_{2}^{*}&t_{2}\end{bmatrix}\succeq \boldsymbol{0},\\
&\quad \quad \ \quad   \quad\quad\quad \   \by= \bx_1 + \boldsymbol{g} \odot \bx_2.
\end{split}
\end{equation}

Define the separation condition of point sources of each channel as
$$\Delta_i =  \min_{k\neq j} \left\vert\tau_{ik} - \tau_{ij}\right\vert,$$
which is the wrapped-around distance on $[0,1]$, and the minimum separation of all channels as $\Delta =\min_i \Delta_i$. Our main theorem is stated below.
\begin{theorem}\label{theorem_main}
Let $M\geq 4$. Assume that $g_{n}=e^{j2\pi\phi_n}$'s are i.i.d. randomly generated from a uniform distribution on the complex unit circle with $\phi_n\sim\mathcal{U}[0,1]$, and that the signs of the coefficients $\boldsymbol{a}_{ik}$'s are i.i.d. generated from a symmetric distribution on the complex unit circle. Provided that the separation $\Delta \ge 1/M$, there exists a numerical constant $C$ such that 
\begin{equation*}
\begin{split}& M\ge C\max\Bigg\{\log^{2}{\left(\frac{M\left( {K_{1}}+ {K_{2}}\right) }{\eta}\right)}, \\ &\max{\{K_{1},K_{2}\}}\log{\left(\frac{K_{1}+K_{2}}{\eta}\right)}\log{\left(\frac{M\left( {K_{1}}+ {K_{2}}\right) }{\eta}\right)}\Bigg\}\\ \end{split}
\end{equation*}
is sufficient to guarantee that $\boldsymbol{x}_{1}^{\star}$ and $\boldsymbol{x}_{2}^{\star}$ are the unique solutions of \eqref{convex_demixing_rewritten} with probability at least $1-\eta$.
\end{theorem}
Theorem~\ref{theorem_main} indicates that as soon as the number of measurements, or alternatively, the bandwidth of the point spread functions, is on the order $M= O(\max(K_1,K_2)\log(K_1+K_2)\log M)$, the proposed convex demixing algorithm recovers the locations of the point sources for each channel exactly, with high probability. This suggests that the performance of the convex demixing algorithm is near optimal in terms of the sample complexity.  

\begin{remark}
The point source separation condition $\Delta\ge 1/M$ is used as a sufficient condition in Theorem~\ref{theorem_main} to guarantee accurate signal demixing. It is implied in \cite{candes2014towards} that a reasonable separation is also necessary to guarantee stable superresolution. Interestingly, no separation between point sources is necessary across channels, as long as their point spread functions are sufficiently incoherent.
\end{remark}

\begin{remark}
Theorem~\ref{theorem_main} assumes $g_{n}$'s are generated with uniformly random phase, which may be reasonable when $g_n$'s can be designed, such as the spreading sequences in multi-user communications. This assumption may be further relaxed. The proof procedure reveals that same results can be obtained as long as $g_{n}$'s satisfy $\mathbb{E}\left[\bar{g}_{n}\right]=\mathbb{E}\left[\bar{g}_{n}^{-1}\right]=0$ and $C_{1}\le\left\vert g_{n}\right\vert\le C_{2}$.
\end{remark}

\begin{remark}
Both $\mbox{sign}\left(a_{1k}\right)$ and $\mbox{sign}\left(a_{2k}\right)$ are required to be randomly generated, which we believe are technical requirements of the proof, and may be removed with finer proof techniques.
\end{remark}

Define the inner product of two vectors as $\langle\boldsymbol{p}, \boldsymbol{x}\rangle=\boldsymbol{x}^{*}\boldsymbol{p}$ and the real-valued inner product as $\langle\boldsymbol{p}, \boldsymbol{x}\rangle_{\mathbb{R}}=\mbox{Re}\left(\boldsymbol{x}^{*}\boldsymbol{p}\right)$. Then the dual norm of $\left\Vert\cdot\right\Vert_{\mathcal{A}}$ can be represented as 
\begin{equation*}
\left\Vert\boldsymbol{p}\right\Vert_{\mathcal{A}}^{\star}=\sup_{\left\Vert\boldsymbol{x}\right\Vert_{\mathcal{A}}\le 1}\ \langle\boldsymbol{p},\boldsymbol{x}\rangle_{\mathbb{R}}=\sup_{\tau\in[0,1)}\left\vert\sum_{n=-2M}^{2M}p_{n}e^{j2\pi n\tau}\right\vert.
\end{equation*}
We have the dual problem of \eqref{convex_demixing_rewritten} written as 
\begin{equation}\label{convex_demixing_dual}
\begin{split}
&\hat{\boldsymbol{p}}=\argmax_{\boldsymbol{p}}\;\langle\boldsymbol{p},\boldsymbol{y}\rangle_{\mathbb{R}},\\
& \quad\quad \mbox{s.t.}\quad \left\Vert\boldsymbol{p}\right\Vert_{\mathcal{A}}^{\star}\le 1,\ \left\Vert\bar{\boldsymbol{g}}\odot\boldsymbol{p}\right\Vert_{\mathcal{A}}^{\star}\le 1,
\end{split}
\end{equation}
which comes from standard Lagrangian calculations. Construct two dual polynomials using the dual solution $\hat{\boldsymbol{p}}$:
$$P\left(\tau\right)=\sum_{n=-2M}^{2M} p_{n}e^{j2\pi n\tau} , \quad  Q\left(\tau\right)=\sum_{n=-2M}^{2M}p_{n}\bar{g}_{n}e^{j2\pi n\tau},$$
then the point source locations $\Upsilon_1$ and $\Upsilon_2$ can be recovered without model order estimation, by identifying the parameter $\tau$ such at $|P(\tau)|=1$, and $|Q(\tau)|=1$, respectively.


\section{Proof Sketch of Main Result}\label{sec::proof_sketch}
In this section, we proceed to sketch the proof of Theorem \ref{theorem_main}. We first provide the optimality conditions using dual polynomials to certify the optimality of the solution of \eqref{convex_demixing_rewritten}. Illuminated by \cite{candes2014towards,tang2014CSoffgrid}, where the dual polynomial is constructed by the squared Fej\'er's kernel, we propose a construction of dual polynomials which are composed of a deterministic term and a random perturbation termed introduced by interference between channels. Finally, the remainder is to show that the constructed dual polynomials satisfy the optimality conditions with high probability when the number of measurements $M$ is large enough.

\subsection{Optimality Conditions using Dual Polynomials}
We can certify the optimality of the primal problem \eqref{convex_demixing_rewritten} using the following proposition.
\begin{prop}\label{dual_certificate}
The solution of \eqref{convex_demixing_rewritten} $\hat{\boldsymbol{x}}_{1}=\boldsymbol{x}_{1}^{\star}$ and $\hat{\boldsymbol{x}}_{2}=\boldsymbol{x}_{2}^{\star}$ is the the unique optimizer if there exists a vector $\boldsymbol{p}$ such that the dual polynomials $P(\tau)$ and $Q(\tau)$ constructed from it
\begin{equation}\label{dual_polynomial}
P\left(\tau\right)=\sum_{n=-2M}^{2M} p_{n}e^{j2\pi n\tau} , \;  Q\left(\tau\right)=\sum_{n=-2M}^{2M}p_{n}\bar{g}_{n}e^{j2\pi n\tau}
\end{equation}
satisfy
\begin{equation} \label{conditions}
\begin{cases}
 P\left(\tau_{1k}\right) = \mbox{sign}\left(a_{1k}\right), & \forall \tau_{1k} \in \Upsilon_{1} \\
  \left\vert P\left(\tau\right)\right\vert < 1, &\forall \tau\notin \Upsilon_{1} \\
  Q\left(\tau_{2k}\right) = \mbox{sign}\left(a_{2k}\right), & \forall \tau_{2k} \in \Upsilon_{2} \\
  \left\vert Q\left(\tau\right)\right\vert < 1, &\forall \tau\notin\Upsilon_{2} \\
\end{cases},
\end{equation}
where the sign should be understood as the complex sign.
\end{prop}
Proposition~\ref{dual_certificate} suggests that $P\left(\tau\right)$ and $Q\left(\tau\right)$ are dual certificates to show $\left\{\bx_{1}^{\star},\bx_{2}^{\star}\right\}$ is the unique primal optimizer. Therefore if we can find a vector $\boldsymbol{p}$ to construct two dual polynomials $P(\tau)$ and $Q(\tau)$ in \eqref{dual_polynomial} satisfying \eqref{conditions}, the proposed algorithm is guaranteed to recover the ground truth. 

\subsection{Dual Certificate Construction}
Our construction of dual polynomials is inspired by \cite{candes2014towards,tang2014CSoffgrid},  based on use of the squared Fej\'er's kernel. However, since the two dual polynomials are coupled together, the construction is more involved. Define the squared Fej\'er's kernel \cite{candes2014towards} as
\begin{equation}\label{func_K}
K\left(\tau\right)=\frac{1}{M}\sum_{n=-2M}^{2M}s_{n}e^{j2\pi n\tau},
\end{equation}
where $s_{n}=\frac{1}{M}\sum_{i=\max\left(n-M,-M\right)}^{\mbox{min}\left(n+M,M\right)}\left(1-\left\vert\frac{i}{M}\right\vert\right)\left(1-\left\vert\frac{n}{M}-\frac{i}{M}\right\vert\right)$. The value of $K\left(\tau\right)$ is nonnegative, attaining the peak at $\tau=0$ and decaying to zero rapidly with the increase of the absolute value of $\tau$.

We define $K_{g}\left(\tau\right)$ and $K_{\bar{g}}\left(\tau\right)$ respectively as 
\begin{equation}\label{func_Kg}
\begin{split}
&K_{g}\left(\tau\right)=\frac{1}{M}\sum_{n=-2M}^{2M}s_{n}{g}_{n}e^{j2\pi n\tau}, \\
&K_{\bar{g}}\left(\tau\right)=\frac{1}{M}\sum_{n=-2M}^{2M}s_{n} \bar{g}_{n} e^{j2\pi n\tau}.
\end{split}
\end{equation}

We then construct the dual polynomials $P\left(\tau\right)$ and $Q\left(\tau\right)$ as
\begin{equation}\label{func_Pf}
\begin{split}
P\left(\tau\right)&=\sum_{k=1}^{K_{1}}\alpha_{1k}K\left(\tau-\tau_{1k}\right)+\sum_{k=1}^{K_{1}}\beta_{1k}K'\left(\tau-\tau_{1k}\right)\\
&+\sum_{k=1}^{K_{2}}\alpha_{2k}K_{g}\left(\tau-\tau_{2k}\right)+\sum_{k=1}^{K_{2}}\beta_{2k}K_{g}'\left(\tau-\tau_{2k}\right),
\end{split}
\end{equation} 
and
\begin{equation}\label{func_Qf}
\begin{split}
Q\left(\tau\right)&=\sum_{k=1}^{K_{1}}\alpha_{1k}K_{\bar{g}}\left(\tau-\tau_{1k}\right)+\sum_{k=1}^{K_{1}}\beta_{1k}K_{\bar{g}}'\left(\tau-\tau_{1k}\right)\\
&+\sum_{k=1}^{K_{2}}\alpha_{2k}K\left(\tau-\tau_{2k}\right)+\sum_{k=1}^{K_{2}}\beta_{2k}K'\left(\tau-\tau_{2k}\right),
\end{split}
\end{equation}
where $\tau_{1k}\in\Upsilon_{1}$ and $\tau_{2k}\in\Upsilon_{2}$. It is straightforward to validate that there exists some corresponding vector $\boldsymbol{p}$ such that \eqref{func_Pf} and \eqref{func_Qf} can be equivalently written as \eqref{dual_polynomial}. Set the coefficients $\boldsymbol{\alpha}_{1}=\left[\alpha_{11},\dots,\alpha_{1K_{1}}\right]^{T}$, $\boldsymbol{\beta}_{1}=\left[\beta_{11},\dots,\beta_{1K_{1}}\right]^{T}$, $\boldsymbol{\alpha}_{2}=\left[\alpha_{21},\dots,\alpha_{2K_{2}}\right]^{T}$ and $\boldsymbol{\beta}_{2}=\left[\beta_{21},\dots,\beta_{2K_{2}}\right]^{T}$ by solving the following equations
\begin{equation}\label{equ_solve_coeff}
\begin{cases}
P\left(\tau_{1k}\right)  =  \mbox{sign}\left(a_{1k}\right), &\quad \tau_{1k}\in\Upsilon_{1}, \\
P'\left(\tau_{1k}\right)  =  0, &\quad \tau_{1k}\in\Upsilon_{1}, \\
Q\left(\tau_{2k}\right)  =  \mbox{sign}\left(a_{2k}\right), &\quad \tau_{2k}\in\Upsilon_{2}, \\
Q'\left(\tau_{2k}\right)  =  0, &\quad \tau_{2k}\in\Upsilon_{2}.
\end{cases}
\end{equation}
The above setting, if exists, immediately satisfies the first and third conditions in \eqref{conditions}. The rest of the proof is then to, under the condition of Theorem~\ref{theorem_main}, guarantee that a solution of \eqref{equ_solve_coeff} exists with high probability, and that when existing, they satisfy the second and forth conditions in \eqref{conditions} with high probability, therefore completing the proof.

\subsection{Validation of Constructed Dual Polynomials}
First we want to show that the solution of \eqref{equ_solve_coeff} exists with high probability. Since the deterministic terms in constructed dual polynomials are well-conditioned if the point source separation $\Delta$ satisfies appropriate condition \cite{tang2014CSoffgrid}, by showing the random perturbations are small, we can have the following proposition to guarantee the invertibility of \eqref{equ_solve_coeff} with high probability when the number of measurements is large enough.
\begin{prop}\label{prop:invertibility}
Assume $M\geq 4$. Let $\delta\in(0,0.6376)$ and $\eta\in(0,1)$, then \eqref{equ_solve_coeff} is invertible with probability at least $1-\eta$ provided that 
\begin{equation}\label{sample_complexity}
M\ge \frac{46}{\delta^{2}}\max\{K_{1},K_{2}\}\log\left(\frac{2\left(K_{1}+K_{2}\right)}{\eta}\right).
\end{equation}
\end{prop}

Once the constructed dual polynomials are fixed with coefficients $\left\{\boldsymbol{\alpha}_{1},\boldsymbol{\beta}_{1},\boldsymbol{\alpha}_{2},\boldsymbol{\beta}_{2}\right\}$ from the solution of \eqref{equ_solve_coeff}, the rest is to verify that $\left\vert P\left(\tau\right)\right\vert < 1$, $\forall \tau\notin \Upsilon_{1}$ and similarly, $\left\vert Q\left(\tau\right)\right\vert < 1, \forall \tau\notin \Upsilon_{2}$. Since the expressions for $P(\tau)$ and $Q(\tau)$ are very similar, it is sufficient to only establish the above for $P(\tau)$.

Denote 
\begin{equation*}
\frac{1}{\sqrt{\left\vert K''\left(0\right)\right\vert}^{l}}\bar{P}^{\left(l\right)}\left(\tau\right)= \mathbb{E}\left[\frac{1}{\sqrt{\left\vert K''\left(0\right)\right\vert}^{l}}P^{\left(l\right)}\left(\tau\right) \right],
\end{equation*}
which is well-bounded \cite{tang2014CSoffgrid, candes2014towards}, where the expectation is with respect to $\boldsymbol{g}$ and $P^{\left(l\right)}\left(\tau\right)$ is the $l$th derivative of $P\left(\tau\right)$. 

While the random perturbation in constructed $P^{\left(l\right)}\left(\tau\right)$ introduced by interference effect can be verified small enough with high probability, we have the following proposition to uniformly bound the distance between $P^{\left(l\right)}\left(\tau\right)$ and $\bar{P}^{\left(l\right)}\left(\tau\right)$. 
\begin{prop}\label{bound_continuous_content}
Suppose $\Delta\ge\frac{1}{M}$. If there exists a numerical constant $C$ such that
\begin{equation*}
\begin{split}
&M\ge C\frac{1}{\epsilon^{2}}\max\Bigg\{\log^{2}{\left(\frac{M\left( {K_{1}}+ {K_{2}}\right)}{\epsilon\eta}\right)},\\
&\max{\{K_{1},K_{2}\}}\log{\left(\frac{K_{1}+K_{2}}{\eta}\right)}\log{\left(\frac{M\left( {K_{1}}+ {K_{2}}\right)}{\epsilon\eta}\right)}\Bigg\},
\end{split}
\end{equation*}
then  
\begin{equation*}
\left\vert\frac{1}{\sqrt{\left\vert K''\left(0\right)\right\vert}^{l}}P^{\left(l\right)}\left(\tau\right) - \frac{1}{\sqrt{\left\vert K''\left(0\right)\right\vert}^{l}}\bar{P}^{\left(l\right)}\left(\tau\right)\right\vert\le\epsilon, 
\end{equation*}
$\forall\tau\in[0,1),\ l=0,1,2,3$, holds with high probability at least $1-\eta$.
\end{prop}

Then applying Bernstein's polynomial inequality \cite{schaeffer1941inequalities} and similar techniques in \cite[Lemma 4.13 and 4.14]{tang2014CSoffgrid}, we have the following proposition.
\begin{prop}\label{bound_value_far}
Suppose $\Delta\ge \frac{1}{M}$. If there exists a numerical constant $C$ such that 
\begin{equation*}
\begin{split}
&M\ge C\max\Bigg\{\log^{2}{\left(\frac{M\left( {K_{1}}+ {K_{2}}\right) }{\eta}\right)},\\
&\max{\{K_{1},K_{2}\}}\log{\left(\frac{K_{1}+K_{2}}{\eta}\right)}\log{\left(\frac{M\left( {K_{1}}+ {K_{2}}\right) }{\eta}\right)}\Bigg\},
\end{split}
\end{equation*}
then 
\begin{equation*}
\left\vert P\left(\tau\right)\right\vert< 1,\ \mbox{for}\;   \tau\in[0,1]\backslash\Upsilon_1,
\end{equation*}
with probability at least $1-\eta$.
\end{prop}
The proof of Theorem \ref{theorem_main} is then complete since we have established that $P(\tau)$ and $Q(\tau)$ constructed in \eqref{func_Pf} and \eqref{func_Qf} are indeed valid dual certificates under the condition of Theorem~\ref{theorem_main}. 


\section{Numerical Experiments}\label{sec::numerical_experiment}
\begin{figure}[h]
\begin{center}
\begin{tabular}{cc} 
\hspace{-0.2in}\includegraphics[width=0.25\textwidth]{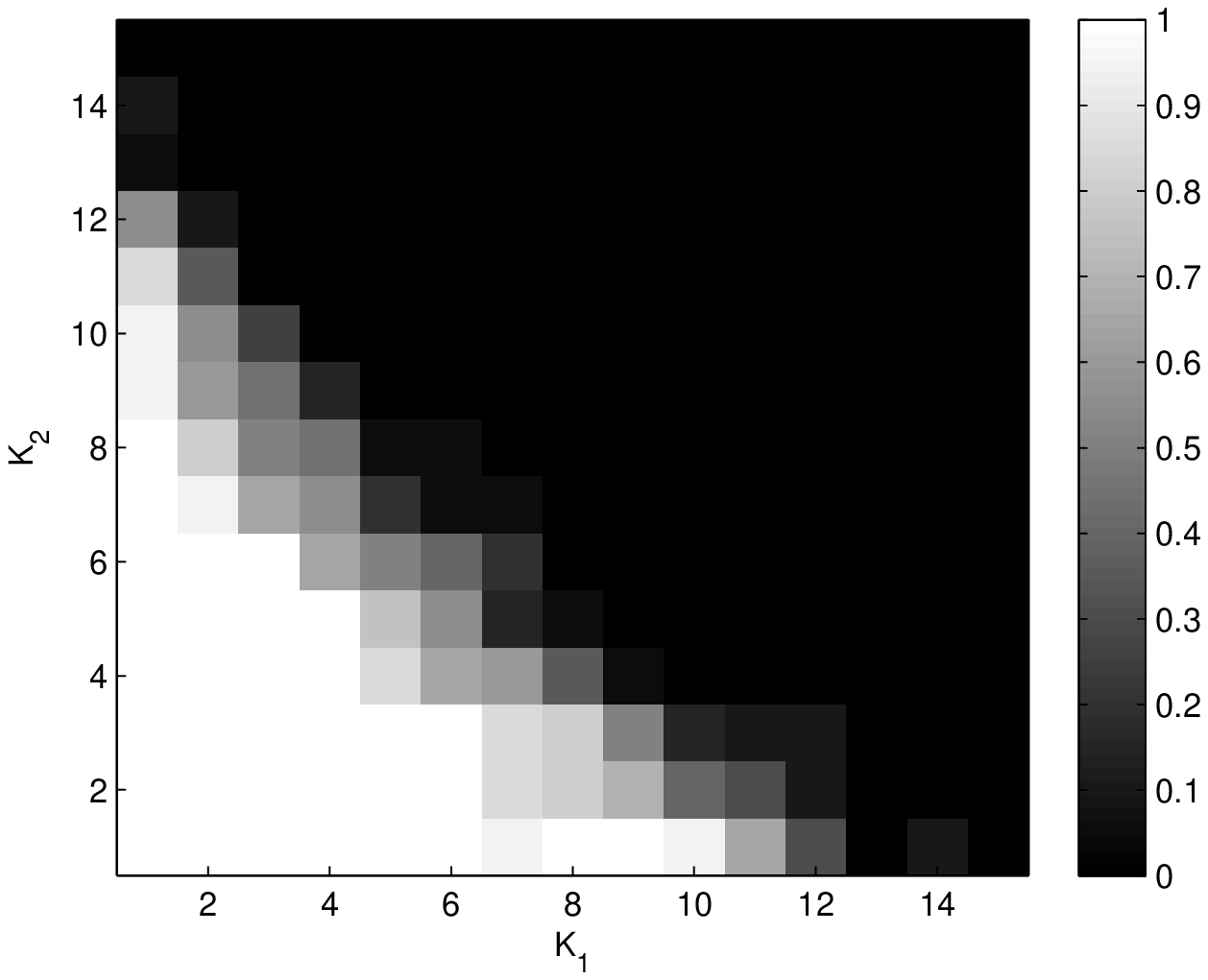} & \hspace{-0.2in}\includegraphics[width=0.25\textwidth]{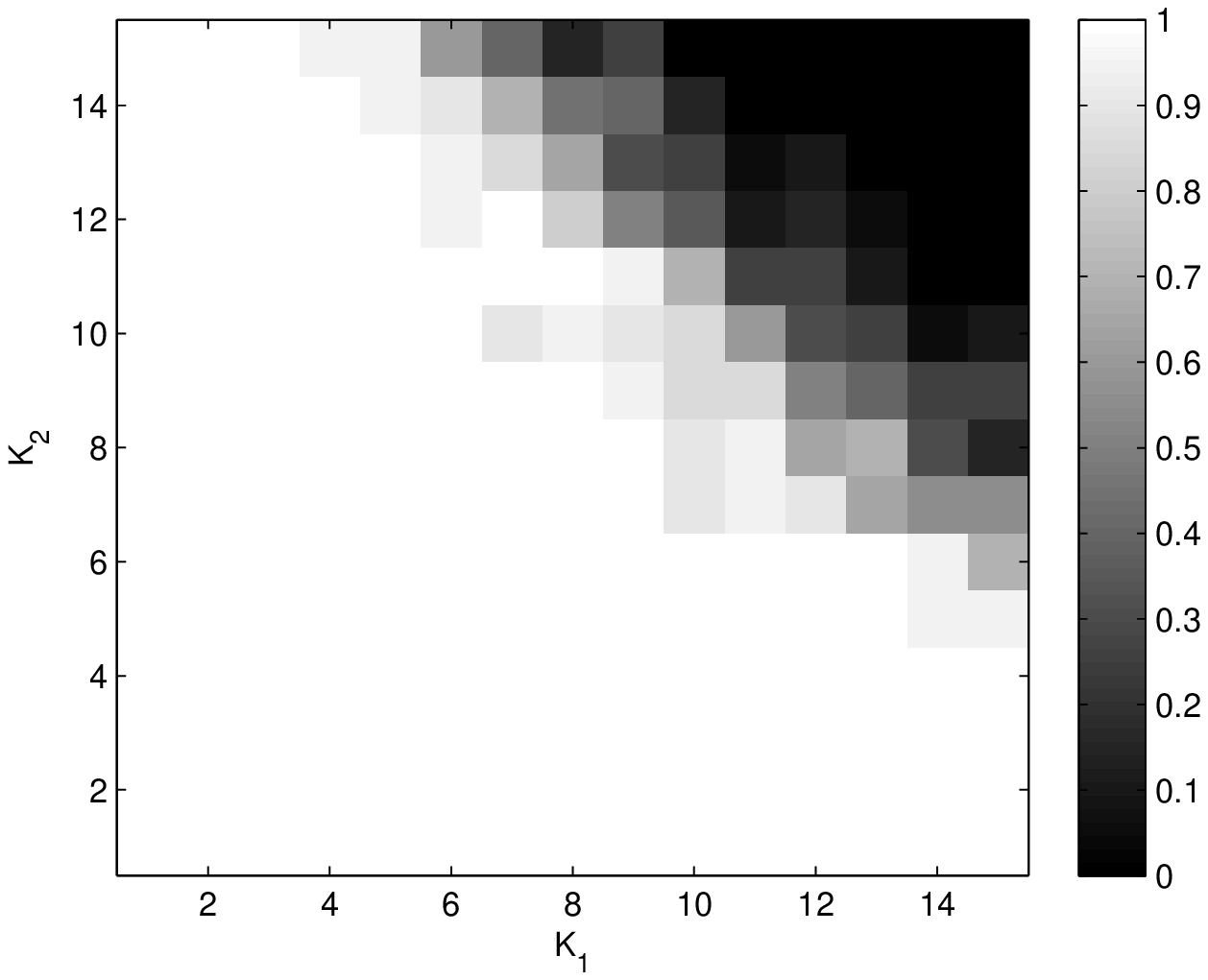} \\
(a) $M = 8$ & (b) $M=16$
\end{tabular}
\end{center}
\caption{Successful rates of the convex demixing algorithm when (a) $M=8$ and (b) $M = 16$.}\label{fig:phase_tran}
\end{figure}

We carry out numerical experiments to validate the performance of the proposed convex demixing algorithm. For a fixed $M$, we vary the spectral sparsity level of the two channels as $K_1$ and $K_2$. For each pair of $(K_1, K_2)$, we first randomly generate a pair of point sources $\Upsilon_{1}$ and $\Upsilon_{2}$ that satisfy a separation condition $\Delta \ge 1/\left(2M\right)$, which is in fact a little smaller than the theoretical constraint, with the coefficients of the point sources i.i.d. drawn from the complex standard Gaussian distribution. For each Monte Carlo trial, we then randomly generate the point spread functions $g_n$'s in the frequency domain with i.i.d. entries drawn uniformly from the complex unit circle, and perform the algorithm by solving \eqref{convex_demixing_sdp} using CVX \cite{grant2008cvx}. The algorithm is considered successful when the Normalized Mean Squared Error (NMSE) satisfies $\sum_{i=1}^{2}\left\Vert\hat{\bx}_{i}-\bx_{i}^{\star}\right\Vert_{2}/\left\Vert\bx_{i}^{\star}\right\Vert_{2}\le 10^{-4}$. Fig.~\ref{fig:phase_tran} shows the success rate of the proposed algorithm over $20$ Monte Carlo trials for each cell, when $M= 8$ in (a) and $M=16$ in (b), respectively.


\section{Conclusions}\label{sec::conclusion}
We propose a convex optimization method based on atomic norm minimization to super-resolve two point source models from the measurements of their superposition, with each convolved with a different low-pass point spread function. It is demonstrated, with high probability, that the point source locations of each channel can be simultaneously determined perfectly, from an order-wise near-optimal number of measurements, under mild conditions. The proposed framework can be extended to handle more than two channels, whose details will be discussed elsewhere.



\section*{Acknowledgment}

This work is supported in part by the Ralph E. Powe Junior Faculty Enhancement Award from the Oak Ridge Associated Universities.



%

\bibliography{draft,bibfileToeplitzPR,bibToeplitz,omp_ref}
\bibliographystyle{IEEEtran}

\end{document}